# Transferability of the light-soaking benefits on silicon heterojunction cells to module

Jean Cattin, Delphine Petri, Jonas Geissbühler, Matthieu Despeisse, Christophe Ballif, Mathieu Boccard

*Abstract* — We investigate the effect of light soaking and forward electric bias treatment on SHJ solar cells and modules, and in particular the influence of the thermal treatment occurring during lamination. A substantial performance increase is observed after electric bias or light soaking, which is shown to be potentially partly reset by the lamination process. This reset is reproduced by annealing the cells with the same thermal budget. A second treatment after lamination again improves performances, and similar final performance is reached independently of the pre-lamination treatment. Therefore, a single treatment after lamination enables maximal module output without any benefit from a cell pre-treatment. Whereas cells react overall better to forward bias, modules show a slightly better response to light soaking.

## I. INTRODUCTION

Silicon heterojunction (SHJ) solar cells [1] are known to increase their efficiency during the first few days of light exposure [2], [3]. This performance improvement was shown to originate from an enhanced defect passivation in the a-Si:H layers when exposed to light, leading to a $V_{oc}$ and FF increase [4]. Wright et al. demonstrated a $+0.7\%_{abs}$ efficiency improvement using a 100 sun exposition with a monochromatic laser, for 30 seconds, reaching a cell temperature of 230°C [5], [6]. This fast process, compatible with industrial production, increased the $V_{oc}$ by 7 mV and the FF by $0.58\%_{abs}$ through a reduction of the series resistance $R_s$ (-0.13 Ωcm$^2$) on Hevel industrial cells. The reduction of active interface defects during light soaking would result from a weakening of the hydrogen ions bonding in the a-Si:H layers, induced by the trapped charges coming from the bulk under illumination [7]. The mobile hydrogen ions would then migrate towards the interface under the presence of an electric field induced by the doped layers and decrease active defect density. The absence of doped layers in the same conditions would result in a lifetime decrease, due to light-induced defects [8], [9]. Light soaking was shown to efficiently increase efficiency also on p-type wafer-based SHJ cells, despite the presence of B-O defects inducing a small subsequent light-induced degradation [10]. However, cells fabricated using p-type wafer generally degrade under light soaking, but this degradation was shown to be significantly limited by a hydrogenation treatment [11]. Degradation on n-type based cells can also occur during light soaking at temperatures above 85°C, but is not inherent to all devices [12].

At the cell level, the beneficial effect of light soaking was shown to saturate after a few days of light exposure [2], [3], reaching an equilibrium between recombination-induced defect creation and healing [13]. Furthermore, the same effect on current-voltage (IV) measurement parameters could be replicated by applying a forward electric current in the dark, which yields an increase of the minority carrier injection in the absorber similarly as under illumination [3], [14]. Some reports show that temperature and light intensity applied during the light soaking treatment have only little influence on the effect kinetics [15]. Other reports show that a high-intensity light soaking using a continuous laser yielding an illumination equivalent to about 80 suns reaches its maximal lifetime improvement potential in approximately one minute [16], which is consistent with recent work on cells [6]. Furthermore, the improvement potential can be enhanced by applying a 200°C annealing before the light soaking treatment [16], tentatively explained by a reduced density of defect states at the interface which enhances hydrogen redistribution [17]. In contrast to many papers highlighting the benefits of the light soaking process for n-type SHJ solar cells, there are also reports showing the possibility for the efficiencies of n-type SHJ solar cells to degrade during light soaking at temperatures above 85°C [12], or even degradation at lower temperatures when a "weak" p-layer (suboptimal thickness or unsufficient doping) is exposed to UV light [18].

Contradictory reports exist on the effect of thermal annealing on the benefits of light soaking. Some reports show mostly stable benefits upon annealing at 200°C for 100 minutes, however small lifetime decrease for ip/ip structures [2]. Other reports show mostly reversible benefits after annealing [16]. The latter study by Bao et.al. shows that cycles of high-intensity light soaking and thermal annealing have a different impact on the different device stacks. Surprisingly, i/c-Si/i improve during light soaking, which contradicts previous studies. Their results suggest a good improvement on n/c-Si/n structures and degradation on p/c-Si/p after repeated cycles, tentatively explained by an increase of the p-layer's doping efficiency [19]. This result, while not in agreement with other studies [4], could be related to the Fermi-level related defects equilibrium, more challenging on the p-side of the device [20]. In all cases, a substantial reset is measured after each annealing at temperatures as low as 160°C, followed by a full recovery and even a small gain during subsequent light soaking, compared to previous cycles, suggesting that a small part of the light soaking benefits is not reversible. However, similar cycles on cells show a fully reversible light soaking effect on the IV properties.

In this study, we investigate the effect of light soaking and forward bias on industrial-grade SHJ cells and modules, and the influence of thermal annealing during the lamination process.

## II. EXPERIMENTAL DETAILS

All the cells used in this study are 6'' SHJ solar cells, purchased on the industrial cell market.

The system used for the light soaking process is a *Solaronix Solixon A-65* degradator, delivering a class A AM1.5G spectrum at 1000 W/m$^2$, with a spatial nonuniformity of class B (IEC 60904-9). The sample temperature is regulated to 50 °C using an aluminium chuck with a water circulation system connected to a *Julabo FC600S* chiller, allowing for a temperature accuracy of ±0.2 °C. The light soaking treatment (LS) is performed on cells under open-circuit (OC) conditions during one week.

The forward electric bias treatment (FWB) consists of applying a forward current density of 40 mA/cm$^2$ (corresponding to the minority carrier injection level reached at $V_{oc}$ under an AM1.5G illumination) in the dark during one week. The power supplies (*TPAE IC-triple*) were connected to cells through aluminium foil contacts and the modules were directly plugged using their metallisation. FWB is applied at ambient temperature without active temperature control and minimal sample heating (~30 °C) coming from dissipation of the applied electric power was observed.

The IV properties of the cells were measured using a Wacom WXS-220S-L2 with a standard deviation of the different IV parameters as follows: Eff: ± 0.05 %$_{abs}$, $V_{oc}$: ± 0.5 mV, FF: ± 0.15 %$_{abs}$, $J_{sc}$: ± 0.04 mA/cm$^2$. The modules were measured using a horizontal Pasan LED flasher with a 162x162 mm$^2$ mask and connected with a four-probe configuration. The standard deviation of the setup is estimated at ±3%, and the reproducibility at ±1%.

The experimental procedure, shown in figure 1, is split into four consecutive IV measurements, separated by three treatment phases. The first treatment (T1) is either reference (storage in the dark at ambient temperature), FWB, or LS. After T1, the cells go either through a lamination process to produce modules or thermal annealing (pseudo-lamination, following a typical lamination duration of about 30 minutes) for cell samples. The second treatment (T2) is either FWB or LS.

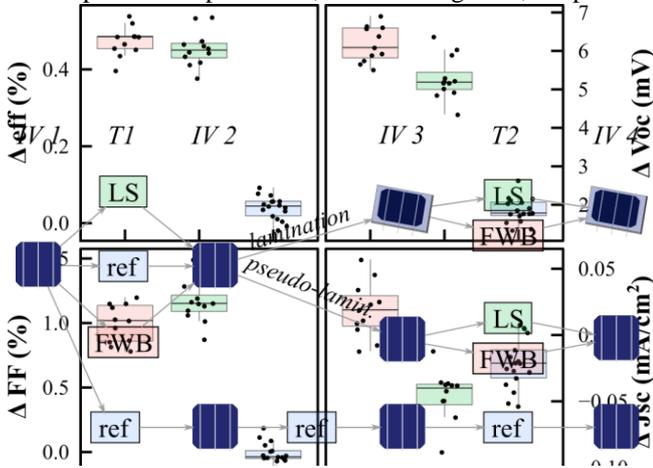

**Figure 1:** Scheme of the experimental procedure. LS: Light soaking, FWB: forward bias, ref: reference.

**Figure 2:** Absolute variations of the IV results caused by the different treatments applied on new cells (T1). LS: Light soaking, FWB: forward bias, ref: reference.

As the effects of the subsequent treatments are expected to affect the $V_{oc}$ and the FF, the cells were sorted by their FF*$V_{oc}$ product and then distributed evenly in the different groups before each treatment phase. This sorting is done to prevent sampling bias. Five cells were selected as global references free from any treatment to calibrate the measurement system along the entire experiment.

## III. RESULTS & DISCUSSION

### A) First treatment: light soaking and forward bias on cells

The initial IV measurements on cells show a narrow distribution of parameters, shown in the Annex in table 1. Figure 2 shows the absolute variations of the IV parameters between the initial measurement and the one after T1. The reference group displays a small $V_{oc}$ gain, most likely caused by a small temperature calibration difference between the two measurement sessions. Both the LS and FWB treatments show an improvement of +0.4%$_{abs}$ in efficiency, a result consistent with previous literature studies. The LS treatment induces a slightly larger FF gain and smaller $V_{oc}$ gain than FWB. Comparisons of IV curves at multiple illuminations [21] reveal that the performance enhancement mostly comes from surface passivation. The small difference between the LS and FWB treatments comes from a slightly smaller passivation gain for LS, along with a small $R_s$ reduction (~0.04 Ω cm). This small difference is currently not explained and could be caused by light exposure of the layers surrounding the wafer, which was shown in previous studies to have an impact on some devices [18]. Additionally, cells exposed to LS show a very small $J_{sc}$ loss. The $R_s$ reduction observed here is smaller than the value reported using fast process under high illumination and temperature [6]. This difference could come from the treatment temperature, not sufficient in this work to modify carrier transport through the interfaces.

### B) Effect of lamination

Part of the cells were then laminated into single-cell modules while another group of cells went through annealing mentioned as *pseudo-lamination*. Figure 3 shows the cell-to-module (CTM) analysis of samples, which represents the absolute IV difference between the cell right before lamination (IV2 on Fig. 1) and the corresponding module after lamination (IV3 on Fig. 1). The first observation is that the reference cells (sample which did not undergo any treatment prior to the lamination) have a better CTM than samples treated prior to lamination, due to $V_{oc}$ and FF differences. This indicates that the gains from the treatments on the cells is partially lost during lamination. Intriguingly, the samples previously exposed to LS (T1 in Fig. 1) undergo a larger FF drop than the ones previously exposed to FWB.

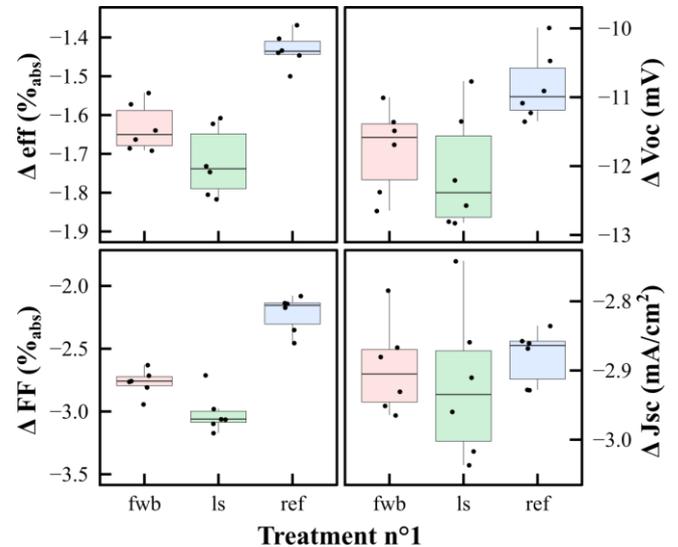

**Figure 3:** Cell-to-module (CTM) performance analysis of the samples laminated into a single cell module, as a function of the pre-lamination treatment (T1). LS: Light soaking, FWB: forward bias, ref: reference.

**Effect of pseudo-lamination**

Cells exposed to a pseudo-lamination (thermal treatment) after the first treatment show a similar behavior to the laminated cells, as shown in figure 4. Whereas the reference cells are insensitive to this annealing treatment, samples previously exposed to a treatment (T1) degrade during annealing, which corresponds well to a partial reset of the benefits from this first treatment. The amplitudes of the IV-parameter variations, shown in Table 2, correspond to those observed during modules lamination. Furthermore, a larger FF reset on samples exposed to LS during T1 compared to the ones exposed to FWB is observed, reproducing the observation made on module samples. Together with the slightly lower FF gain from the FWB compared to LS, this suggests that the LS treatment actually had a stronger effect than FWB, despite a similar injection level in both treatments. This could be due to the lower temperature during FWB compared to LS, which would deserve further investigations [12].

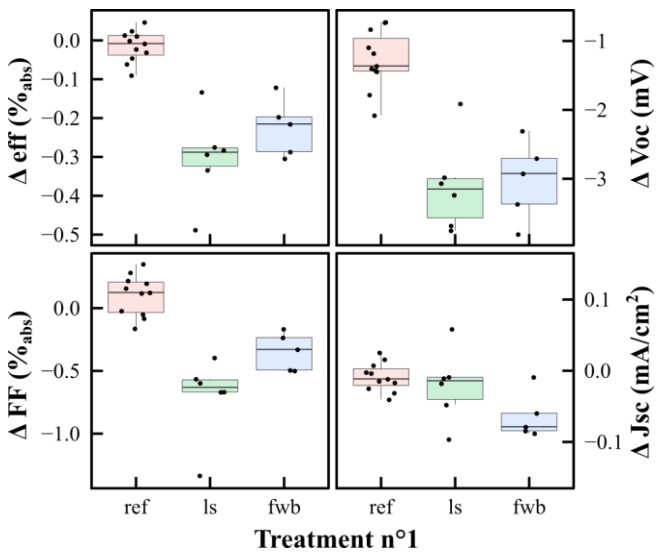

**Figure 4:** Absolute variations of the IV results caused by a pseudo-lamination (thermal annealing) treatment, as a function of the first treatment (T1). LS: Light soaking, FWB: forward bias, ref: reference.

**On the partial IV properties reset upon annealing**

According to these results, the structural changes induced by carrier injection are partially reversible with temperature. This observation contrasts with some studies, where the LS benefits were either stable upon annealing [2] or almost fully reversible [16]. Note that the latter study performed the annealing until the lifetime was stable, while we applied a defined annealing duration. Kobayashi et.al. show an asymmetry between the *i-n* and the *i-p* sides, as symmetric *ip-ip* samples were shown to lose part of the LS benefits on minority carrier lifetime during annealing, and were recovering during further LS [2]. It is believed that this difference is caused by Fermi level dependant H2 bond rupture [20], [22]. This result suggests that the (p) side is responsible for the partial reset of the cells properties in our data. Tentatively, our results could be explained by a pool of defects, passivated by mobile hydrogen ions during LS, and following a distribution of thermal annealing energies, similarly to light-induced defects in a-Si:H [23]. It would mean that our lamination process reactivates only the defects with low activation energy, but there are certainly conditions for which a larger part of the LS benefits are reset (such as longer annealing time or higher temperature). For example, Hata et.al. show that light-induced defects in a-Si:H is fully reversible by annealing for 1 hour at 220°C [24]. There may also be permanent light-induced structural changes, as Stutzmann et al. observed some irreversible changes after multiple LS-annealing cycles [7]. In this case, further experiments probing the curing kinetics during annealing at different temperatures could provide further insights on the nature of the defects.

C) **Second treatment: light soaking and forward bias on modules**

Figure 5 shows absolute variations induced by the second treatment on modules, as a function of the first treatment. Table 2 shows the results of a linear regression analysis of the same results, indicating the magnitude of each treatment's effect on the IV parameters, as well as the statistical relevance of each result. The module $V_{oc}$ is lowered by the second treatment, which could be a module degradation due ot the treatment, or independent of the treatment, or even simple fluctuations in the temperature calibration between the two days of measurement. (No reference module was kept during T2 due to sample availability.) Such drop still enables to analyse the differences between the two treatments.

The samples not previously exposed to a first treatment (T1) are those benefiting the most from the second treatment (T2), which is consistent with the hypothesis that the other samples retained part of the gains obtained during the first treatment. Also, the cells exposed to LS during the first treatment improve slightly more than the cells exposed to FWB.

During the second treatment, the FF and $V_{oc}$ of the modules react better to LS than to FWB. This result is surprising, as all the cells react better to FWB during both treatments 1 and 2. This difference is not yet understood but could be caused by the different treatment temperatures. The small electrical advantage of LS over FWB is compensated by a small but significant Isc reduction during LS. No yellowing was observed on the modules, and this small effect should be experimentally confirmed with a larger amount of samples and dedicated analysis.

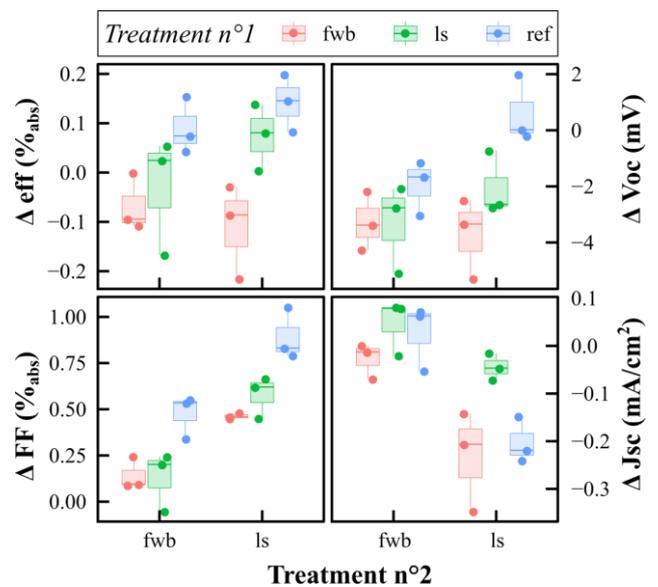

**Figure 5:** Absolute variations of the IV results caused by a FWB or LS treatment on single-cell modules, as a function of the pre-lamination treatment (T1) applied on cells. LS: Light soaking, FWB: forward bias, ref: reference.

### D) Second treatment: light soaking and forward bias on cells

Figure 6 shows the absolute variations induced by the second treatment on cells, as a function of the first treatment. Table 2 shows a linear regression analysis of the same results, indicating the magnitude of each treatment's effect on the IV parameters, as well as the statistical relevance of each result. Similarly to the module case, the cells not exposed to treatment before the pseudo-lamination improve more during the treatment n°2 than cells previously exposed to LS or FWB. A small trend suggests that, like modules, cells exposed to LS during T1 improve their $V_{oc}$ slightly more during the second treatment than cells exposed to FWB, but a larger sample would be needed to confirm this trend.

The second treatment improves the FF similarly during LS and FWB. However, contrarily to the modules, the $V_{oc}$ improves $2 \pm 0.3$ mV more during T2 for FWB than for LS. This preference for FWB is even larger than the one observed during T1. This could be due to a slight loss of selectivity due to light exposure, as previously reported for some architectures [18].

Some variations are also detected on the $J_{sc}$. Similarly to the modules, a small drop of current is observed after LS, only for samples non-previously exposed to LS during T1. This would result in a 0.024 A current drop in the modules presented above, which is smaller than the current drop measured in modules. Such a current drop was not measured in our previous LS experiment on cells and is close to the measurement uncertainty, so should be taken with caution.

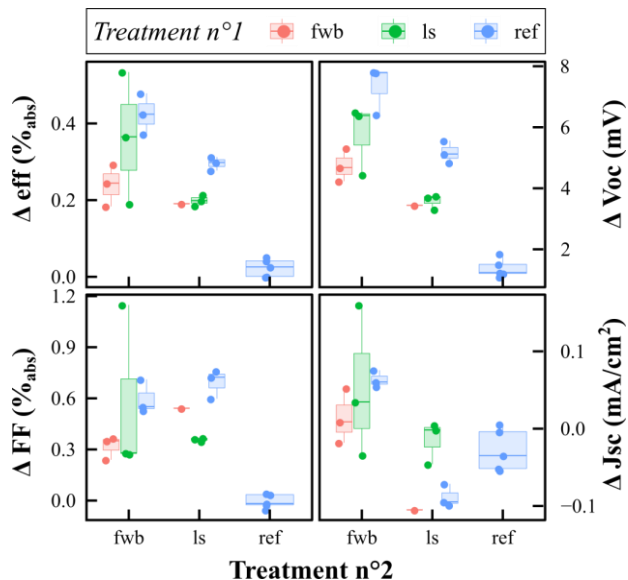

**Figure 6:** Absolute variations of the IV results caused by a FWB or LS treatment on cells, as a function of the treatment applied before thermal annealing (T1). LS: Light soaking, FWB: forward bias, ref: reference.

### E) Overall variation: Where should an LS or FWB treatment be applied?

After detailing each experimental iterations, this paragraph focuses on the overall IV parameters variations, between the initial cell state and the final cell and module state, i.e. after treatment 1 – (pseudo) lamination – treatment 2.

The conclusions differ a bit between cells and modules. For cells, as shown in Figure 7, the treatment n°1 has no significant effect on the overall variation, suggesting that the second treatment is sufficient to reach full benefits. As previously observed on cells, the FWB has a small edge over the LS treatment, with a $2.4 \pm 0.3$ mV larger $V_{oc}$ gain, and no $J_{sc}$ loss.

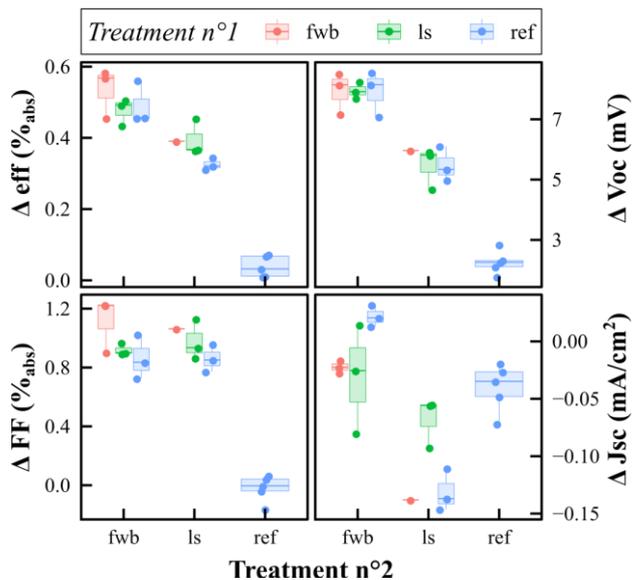

**Figure 7:** Absolute variations of the IV results between the initial and final cells state, as a function of the T1 and T2, applied before and after the pseudo-lamination respectively. LS: Light soaking, FWB: forward bias, ref: reference.

The results for modules are shown in Figure 8. Noticeable differences from the cells follow: Firstly, T1 does influence the final $V_{oc}$ (and possibly FF), but only if FWB is applied as T2. Secondly, $V_{oc}$ and FF benefit more from LS than FWB as T2, as detailed in the previous paragraph. Thirdly, the LS treatment on modules induces a $J_{sc}$ loss, compensating for the FF and $V_{oc}$ advantage.

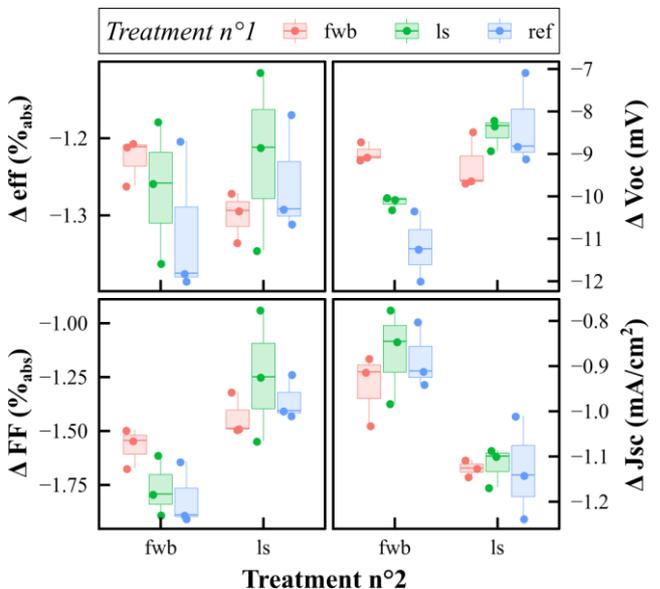

**Figure 8:** Absolute variations of the IV results between the initial cells and final module state, as a function of the T1 and T2, applied before and after the lamination respectively.

These differences between cells and modules are yet to be fully understood and can be separated into two phenomena. Whereas the $J_{sc}$ of cells slightly degraded under LS, stronger degradation is seen for modules which suggest an additional optical effect from the encapsulation. This should be resolved to fully benefit from LS on modules. The differences in $V_{oc}$ and FF could possibly stem from the differences in temperature during LS and FWB, and between cell and module states. A dedicated experiment with more samples or a more accurate measurement system would be needed to identify these small differences accurately. Also, keeping a reference module would be helpful to better track absolute variations caused by the different treatments.

## IV. CONCLUSION

We investigate the effect of a light soaking (LS) and forward electrical bias (FWB) treatment on cells and modules before and after a lamination (for modules) or an equivalent thermal treatment (for cells). The first treatment shows an improvement of the cell performance for both LS and FWB, as commonly reported in the literature for SHJ cells. We report a partial reset of the IV properties during the module lamination, which could be replicated on cells with an annealing mimicking the thermal treatment applied during lamination.

The second treatment significantly improves all samples again, with noticeable differences between the cells and modules. Whereas the cells react overall better to FWB thanks to a larger $V_{oc}$ gain, the modules show a better $V_{oc}$ and FF after LS, but a lower $J_{sc}$. The causes of these small differences deserve further studies.

Therefore, to maximise the watt output of a module, applying a single treatment after lamination appears as the best choice.

## ACKNOWLEDGEMENTS
The authors would like to thank C. Bucher, N. Fürst, S. Pittet and T. Auderset for technical support, J. Levrat for his help with data formatting and A. Faes for his help on project organization.

## V. ANNEXES

**Table 1:** Absolute IV parameters and corresponding standard deviation (SD) of samples along the experiment, as a function of treatment history. T1: Treatment n°1, T2: Treatment n°2, Mod.: module

| Status | T1 | State | T2 | Eff (%) | S.D. (%) | Voc (mV) | S.D. (mV) | FF (%) | S.D. (%) | Jsc (mA/cm²) | S.D. (mA/cm²) |
|---|---|---|---|---|---|---|---|---|---|---|---|
| Init | - | - | - | 22.01 | 0.04 | 734.1 | 2.4 | 79.7 | 0.3 | 37.6 | 0.1 |
| T1 | ref | - | - | 22.05 | 0.04 | 735.3 | 2.6 | 79.7 | 0.3 | 37.6 | 0.1 |
| T1 | LS | - | - | 22.47 | 0.04 | 739.4 | 1.9 | 81.0 | 0.3 | 37.5 | 0.1 |
| T1 | FWB | - | - | 22.48 | 0.04 | 741.3 | 2.0 | 80.7 | 0.3 | 37.6 | 0.1 |
| Lamin. | ref | Mod. | - | 20.68 | 0.08 | 724.8 | 2.0 | 77.8 | 0.6 | 34.7 | 0.2 |
| Lamin. | LS | Mod. | - | 20.75 | 0.07 | 725.2 | 2.9 | 78.2 | 0.3 | 34.6 | 0.1 |
| Lamin. | FWB | Mod. | - | 20.79 | 0.08 | 727.2 | 2.8 | 78.1 | 0.3 | 34.6 | 0.1 |
| Anneal. | ref | Cell | - | 22.03 | 0.08 | 733.7 | 2.7 | 79.8 | 0.5 | 37.6 | 0.1 |
| Anneal. | LS | Cell | - | 22.19 | 0.10 | 737.1 | 2.3 | 80.3 | 0.2 | 37.5 | 0.2 |
| Anneal. | FWB | Cell | - | 22.27 | 0.06 | 739.0 | 1.9 | 80.4 | 0.2 | 37.5 | 0.2 |
| T2 | ref | Mod. | LS | 20.77 | 0.05 | 724.5 | 2.1 | 78.5 | 0.2 | 34.6 | 0.2 |
| T2 | ref | Mod. | FWB | 20.71 | 0.08 | 724.4 | 2.1 | 77.9 | 0.5 | 34.7 | 0.1 |
| T2 | LS | Mod. | LS | 20.79 | 0.09 | 725.5 | 3.1 | 78.5 | 0.3 | 34.5 | 0.1 |
| T2 | LS | Mod. | FWB | 20.73 | 0.08 | 722.1 | 1.4 | 78.3 | 0.3 | 34.7 | 0.1 |
| T2 | FWB | Mod. | LS | 20.69 | 0.03 | 725.2 | 3.0 | 78.2 | 0.2 | 34.5 | 0.1 |
| T2 | FWB | Mod. | FWB | 20.80 | 0.06 | 725.6 | 0.4 | 78.3 | 0.3 | 34.6 | 0.1 |
| T2 | ref | Cell | LS | 22.38 | 0.07 | 737.5 | 0.8 | 80.8 | 0.2 | 37.5 | 0.0 |
| T2 | ref | Cell | FWB | 22.52 | 0.06 | 742.0 | 2.0 | 80.8 | 0.2 | 37.6 | 0.1 |
| T2 | LS | Cell | LS | 22.42 | 0.01 | 739.6 | 2.3 | 80.7 | 0.2 | 37.5 | 0.1 |
| T2 | LS | Cell | FWB | 22.53 | 0.05 | 744.0 | 0.7 | 80.7 | 0.5 | 37.5 | 0.1 |
| T2 | FWB | Cell | LS | 22.40 | NA | 742.1 | NA | 80.9 | NA | 37.3 | NA |
| T2 | FWB | Cell | FWB | 22.53 | 0.01 | 744.8 | 1.1 | 80.7 | 0.3 | 37.5 | 0.1 |

**Table 2:** Results of the linear regression of the absolute IV parameters variations at different experimental stages, in the form of predictor values and corresponding standard error. The analysis was applied at each stage without interactions or higher orders. The coefficients with a P-value below 5% (thus considered as significant) are marked in green.

| Status | Coef. | Eff (%) | S.E. (%) | Voc (mV) | S.E. (mV) | FF (%) | S.E. (%) | Jsc (mA/cm²) | S.E. (mA/cm²) |
|---|---|---|---|---|---|---|---|---|---|
| T1 | Intercept (T1 : ref) | 0.04 | 0.01 | 1.9 | 0.1 | 0.0 | 0.0 | 0.0 | 0.0 |
| T1 | T1: LS | 0.41 | 0.01 | 3.4 | 0.2 | 1.2 | 0.1 | 0.0 | 0.0 |
| T1 | T1: FWB | 0.43 | 0.01 | 4.3 | 0.2 | 1.0 | 0.1 | 0.0 | 0.0 |
| Lamin. | Intercept (T1 : ref) | -1.43 | 0.03 | -10.8 | 0.3 | -2.2 | 0.1 | -2.9 | 0.0 |
| Lamin. | T1: LS | -0.29 | 0.04 | -1.2 | 0.4 | -0.8 | 0.1 | 0.0 | 0.0 |
| Lamin. | T1: FWB | -0.20 | 0.04 | -0.9 | 0.4 | -0.5 | 0.1 | 0.0 | 0.0 |
| Pseudo lamin. | Intercept (T1: ref) | -0.01 | 0.02 | -1.3 | 0.2 | 0.1 | 0.1 | 0.0 | 0.0 |
| Pseudo lamin. | T1: LS | -0.29 | 0.04 | -1.8 | 0.3 | -0.8 | 0.1 | 0.0 | 0.0 |
| Pseudo lamin. | T1: FWB | -0.21 | 0.04 | -1.7 | 0.3 | -0.4 | 0.1 | -0.1 | 0.0 |
| T2 mod. | Intercept (T1:ref, T2:FWB) | 0.10 | 0.04 | -1.2 | 0.6 | 0.5 | 0.1 | 0.0 | 0.0 |
| T2 mod. | T1: LS | -0.09 | 0.05 | -2.0 | 0.8 | -0.3 | 0.1 | 0.0 | 0.0 |
| T2 mod. | T1: FWB | -0.21 | 0.05 | -2.8 | 0.8 | -0.4 | 0.1 | 0.0 | 0.0 |
| T2 mod. | T2: LS | 0.04 | 0.04 | 1.1 | 0.6 | 0.4 | 0.1 | 0.0 | 0.0 |